\documentclass[a4paper]{article}
\usepackage{INTERSPEECH2021}
\usepackage{amsmath}
\usepackage{color}
\usepackage{amssymb, bm, mathtools,array}
\usepackage{dsfont}
\usepackage{verbatim}
\usepackage{lipsum}

\usepackage[title]{appendix}
\usepackage{subfigure}

\usepackage{enumitem}
\setlist[description]{style=nextline}

\usepackage{pifont}

\usepackage{multirow}
\usepackage{multicol}
\usepackage{bm}

\title{An Initial Investigation for Detecting Partially Spoofed Audio}
\name{Lin Zhang$^{1,2}$, Xin Wang$^1$, Erica Cooper$^1$, Junichi Yamagishi$^{1,2}$, Jose Patino$^3$, Nicholas Evans$^3$}
\address{
  $^1$National Institute of Informatics, Japan $^2$SOKENDAI, Japan\\
  $^3$Digital Security Department, EURECOM, France}
\email{\{zhanglin, wangxin, ecooper, jyamagis\}@nii.ac.jp, \{patino, evans\}@eurecom.fr}

\begin{document}

\maketitle
\begin{abstract}

All existing databases of spoofed speech contain attack data that is spoofed in its entirety.  In practice, it is entirely plausible that successful attacks can be mounted with utterances that are only partially spoofed.
By definition, partially-spoofed utterances contain a mix of both spoofed and bona fide segments, which will likely degrade the performance of countermeasures trained with entirely spoofed utterances.  This hypothesis raises the obvious question: \textit{‘Can we detect partially-spoofed audio?’} This paper introduces a new database of partially-spoofed data, named PartialSpoof, to help address this question. This new database enables us to investigate and compare the performance of countermeasures on both utterance- and segmental- level labels. 
Experimental results using the utterance-level labels reveal that the reliability of countermeasures trained to detect fully-spoofed data is found to degrade substantially when tested with partially-spoofed data, whereas training on partially-spoofed data performs reliably in the case of both fully- and partially-spoofed utterances. Additional experiments using segmental-level labels show that spotting injected spoofed segments included in an utterance is a much more challenging task even if the latest countermeasure models are used.
\end{abstract}

\section{Introduction}
To address the problem of presentation attacks \cite{jain2006biometrics} on automatic speaker verification (ASV) systems, the ASVspoof challenge, which aims to promote the study and development of spoofing countermeasures (CMs), has been held biennially since 2015.  Previous challenges \cite{Wu2014, Kinnunen2017, Nautsch2021} focused on different types of spoofing attacks at the utterance level, including the logical access (LA) scenario (i.e.,\ speech synthesis and voice conversion attacks) and the physical access (PA) scenario (i.e.,\ replay attacks). Year by year, various types of high-performance CMs have been proposed. In the latest competition, ASVspoof 2019~\cite{Nautsch2021}, several discriminative (Light CNN~\cite{lavrentyeva2019stc}, ResNet~\cite{Chen2020Odyssey}, Wave-U-Net~\cite{Chettri2019}, RawNet~\cite{tak2020end}) and generative (GMM-UBM) models were proposed which achieved promising results.

When we use spoofing CMs for more general situations beyond presentation attacks against ASV, such as audio deepfake detection, it is obvious that the assumption that attacks will consist of entirely-spoofed utterances does not always hold -- speech synthesis and voice conversion technologies may be used to generate only a part of an utterance, and such spoofed segment(s) may be injected into a bona fide utterance. For example, attackers may use speech synthesis to replace specific phrases that they want to manipulate. This leads us to consider the following new scientific questions: \textit{`Can the latest CMs discriminate such partially-spoofed audio from bona fide audio reliably? Can we construct a new CM model that can detect such injected partially-spoofed segments?'} To investigate this as-yet neglected attack scenario deeply, we first collected a new partially-spoofed database, named ``PartialSpoof'', and used it to evaluate existing CMs in terms of both of utterance- and segmental-level detection of partially spoofed audio. 

As for the existing CMs, we use discriminatively trained Light Convolutional Neural Networks (LCNN) \cite{Nautsch2021, lavrentyeva2019stc} using the P2SGrad loss function \cite{wang2021comparative} as it is reported as the best single model. We train them using either utterance- or segmental-level labels and investigate their detection performance. We also revise the LCNN architecture slightly so that LCNN-based CMs can handle temporal information better when only small fractions of an utterance may be spoofed. 

For experiments on utterance-level detection of partially spoofed audio, we have compared CMs trained to detect fully-spoofed data of the ASVspoof 2019 dataset with equivalent CMs trained on partially-spoofed data from the new PartialSpoof database. For experiments on segmental-level detection of partially spoofed audio, we demonstrate the performance of CMs using the segmetal-level labels instead of utterance-level labels. 

This paper is structured as follows: Section 2 overviews the construction process of the PartialSpoof Database. Section 3 briefly introduces the CMs used for our investigations. Section 4 shows experimental conditions and results, and our findings are summarized in Section 5. 

\section{PartialSpoof Database}\label{sec:databse}

We built the PartialSpoof database\footnote{{Database: https://zenodo.org/record/4817532\#.YLd8Yi2l1hF}} based on the ASVspoof 2019 LA database \cite{Wang2020data} since the latter covers 17 types of spoofed data produced by advanced speech synthesizers, voice converters, and hybrids. We used the same set of bona fide data from the ASVspoof 2019 LA database but created partially-spoofed audio from the ASVspoof 2019 LA data by following the steps below:

\noindent 
\textbf{Step 1:} We ran three types of publicly-available voice activity detection (VAD) algorithms: an energy-based VAD from Kaldi \cite{povey2011kaldi}, another energy-based VAD \cite{kinnunen2010overview}, and an LSTM-based VAD from Pyannote \cite{Lavechin-sad-dihard,pyannote}. 
Then, we used their voting results to decide the boundaries of speech segments in order to reduce VAD errors. 
Specifically, we considered a segment to be speech if it was detected by at least two out of three VAD systems\footnote{This voting of VAD systems can achieve an 11.98\% detection error rate \cite{pyannote.metrics} when evaluated on TIMIT \cite{Zue1990timit}.}. 

\noindent 
\textbf{Step 2:} According to the boundaries determined by the above VAD results, we replaced a randomly chosen segment from a bona fide utterance with a spoofed segment.
We also considered the opposite direction for segment replacement, i.e., randomly substituting a spoofed segment into a bona fide segment. 
There are a few restrictions: 1) The same inserted segment cannot appear more than once in a given carrier utterance; 2) The segment to be inserted must be close in duration to the original segment it replaces. 

\noindent 
\textbf{Step 3:} To avoid potential artifacts when fusing the waveform of the inserted segment into the carrier audio file, we computed the time-domain cross correlation between the replacement segment and its adjacent segments to find the best fusion point. The fusion was then conducted through waveform overlap-add after waveform amplitude normalization using SV56 \cite{sv56}. 
We kept 50\% of the non-speech part in the head and tail of the segments so that overlap-add only happened in such non-speech parts without modifying the speech segment. 

\noindent 
\textbf{Step 4:} We labeled each segment in the newly-created audio as \textit{bona fide} if the segment is originally from bona fide audio, otherwise as \textit{spoofed}. The utterance-level label of a partially-spoofed utterance is of course \textit{spoofed}.

\noindent 
\textbf{Step 5:} We repeated Step 2-4 until we obtained the same number of spoofed trials as the ASVspoof 2019 database. 

Thus, numbers of spoofed trials \footnote{{Samples: \scriptsize{https://nii-yamagishilab.github.io/zlin-demo/IS2021/index.html}}} in the train, dev, and eval sets are the same as those of the ASVspoof 2019 database.

\section{Spoofing Countermeasures} \label{sec:cm}

To determine whether the latest CMs can discriminate partially-spoofed utterances from bona fide ones, we built a series of CMs based on the top single LCNN model in the ASVspoof 2019 LA task \cite{Nautsch2021, lavrentyeva2019stc} but with a few enhancements \cite{wang2021comparative}. Here we explain how we train them for utterance- and segmental-level detection, respectively.  

\subsection{CM training for utterance-level detection}
\label{sec:utterance}

The LCNN used by the top single CM in the ASVspoof 2019 LA scenario only accepts fixed-size inputs~\cite{wu2018light}. The CM hence operates on trimmed or padded input speech~\cite{lavrentyeva2019stc} and is trained to predict an utterance-level score. 
This is unsuitable for the partially-spoofed scenario because the intervals of interest, e.g.,\ the replaced, substituted, or spoofed intervals, might well be among the trimmed segments. 

\begin{figure}[t]
  \centering
    \includegraphics[trim=20 300 20 80, clip, width=0.95\linewidth]{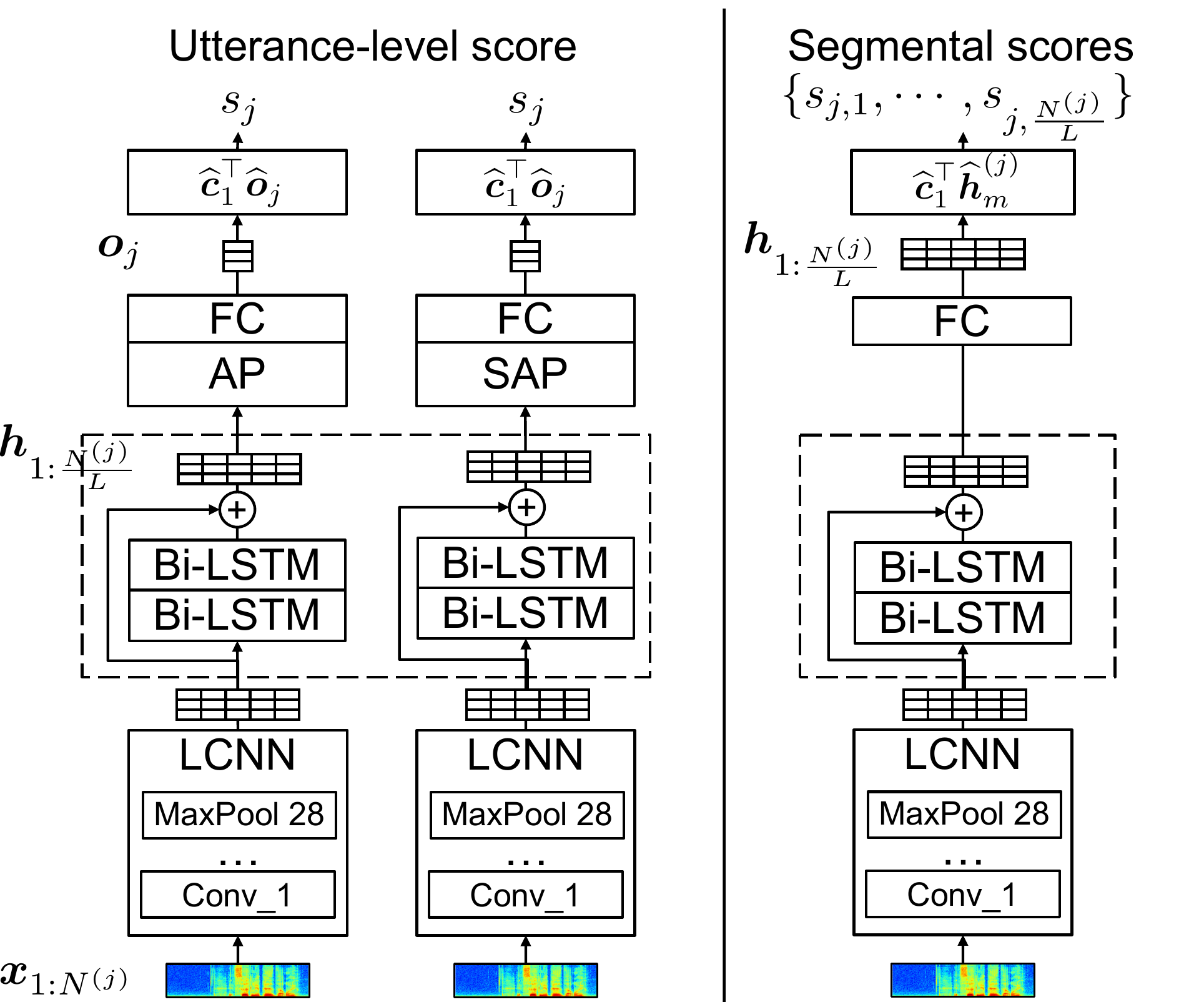}
  \caption{Enhanced LCNNs for conventional and partially-spoofed scenarios. 
  The LCNN part is identical to that in \cite{lavrentyeva2019stc} (from layer {Conv\_1} to  {MaxPool\_28}). FC denotes a fully-connected layer. AP and SAP denote average and self-attentive pooling, respectively. The Bi-LSTM block in the dashed frame is optional.
  }
  \label{fig:spfcon_nn}
  \vspace{-3mm}
\end{figure}

We therefore need to slightly enhance the LCNN with temporal pooling strategies to better process variable-length speech inputs \cite{wang2021comparative}.
Let $\boldsymbol{x}_{1:N^{(j)}} \equiv (\boldsymbol{x}_1, \cdots, \boldsymbol{x}_{N^{(j)}})\in\mathbb{R}^{N^{(j)}\times{D}}$ be the input feature sequence of the $j$-th trial with $N^{(j)}$ frames, where $\boldsymbol{x}_n$ is the feature for the $n$-th frame.
Instead of trimming or padding $\boldsymbol{x}_{1:N^{(j)}}$ to a fixed shape, we let the LCNN transform $\boldsymbol{x}_{1:N^{(j)}}$ into $\boldsymbol{h}_{1:N^{(j)}/L}=(\boldsymbol{h}_{1}^{(j)},\cdots, \boldsymbol{h}_{N^{(j)}/L}^{(j)})$,
where $L$ is decided by the convolution stride.
Then, we pool an utterance-level vector $\boldsymbol{o}_j = \sum_{m=1}^{N^{(j)}/L}w_m^{(j)}\boldsymbol{h}_{m}^{(j)}$ and use it for scoring. The pooling weight $w_m^{(j)}$ can be computed using a self-attentive pooling (SAP) strategy~\cite{Zhu2018} or be uniform, i.e., the average pooling (AP) strategy. The enhanced LCNNs are illustrated in Figure~\ref{fig:spfcon_nn}.

We may optionally use a bi-directional LSTM to process $\boldsymbol{h}_{1:N^{(j)}/L}$ before pooling. 
The reason is that convolution in an LCNN has a fixed receptive field, and each $\boldsymbol{h}_{m}$ covers only a fixed number of input frames. A pure LCNN hence neglects the temporal change that may be useful in the partially-spoofed scenario.
In practice, we add a skip-connection over the Bi-LSTM layer(s) to stabilize the training process. 
This optional Bi-LSTM block is illustrated in Figure~\ref{fig:spfcon_nn}.

\label{sec:utt_utt}
The enhanced LCNNs are trained given pairs of input audio and utterance-level labels $\{\boldsymbol{x}_{1:N^{(j)}}, y_j\}_{j=1}^{|\mathcal{D}|}$, where $|\mathcal{D}|$ is the size of the training data set $\mathcal{D}$.
In this paper, we used a new loss function called \textit{MSE for P2SGrad} since it was found to be more efficient than cross-entropy with variants of softmax on this task \cite{wang2021comparative}.
The loss is computed as
\begin{equation}
\label{eql:utt_score}
\mathcal{L}^{(\text{p2s})} = \frac{1}{|\mathcal{D}|}\sum_{j=1}^{|\mathcal{D}|}\sum_{k=1}^{C} (\cos\theta_{j,k} - \mathds{1}(y_j=k)) ^ 2,
\end{equation}
where $\mathds{1}(\cdot)$ is an indicator function, $C$ is the number of target classes, and $\cos\theta_{j, k} = \widehat{\boldsymbol{c}}_k^\top\widehat{\boldsymbol{o}}_j$ is the cosine distance between the length-normed vector $\widehat{\boldsymbol{o}}_j = \boldsymbol{o}_j / || \boldsymbol{o}_j ||$ and the class vector $\widehat{\boldsymbol{c}}_k$ for the $k$-th target class.
In this paper, we set $C=2$ and use $k=1$ and $k=2$ to denote \textit{bona fide} and \textit{spoofed}, respectively.
During inference, the model uses $s_j = \cos\theta_{j,1}=\widehat{\boldsymbol{c}}_1^\top\widehat{\boldsymbol{o}}_j$ as the utterance-level score for the $j$-th test trial.

\subsection{Deriving segmental scores from an utterance-level score}
\label{sec:utt_seg}

The enhanced LCNNs produce one score per trial. If they are used for segmental-level detection rather than utterance-level detection, then we need to decompose the utterance-level score and derive a score for each segment of the input trial. Here we describe our decomposition procedures.

Suppose the LCNN has conducted $\boldsymbol{x}_{1:N^{(j)}}\mapsto\boldsymbol{h}_{1:M_j}\mapsto\boldsymbol{o}_j = \sum_{m=1}^{M_j}w_m^{(j)}\boldsymbol{h}_{m}^{(j)}$,  
where $M_j=\frac{N^{(j)}}{L}$. 
With the utterance-level score $s_j = \cos\theta_{j,1}= \widehat{\boldsymbol{c}}_1^{\top} \frac{\boldsymbol{o}_j}{||\boldsymbol{o}_j||} $, we get
\begin{equation} 
s_j= \widehat{\boldsymbol{c}}_1^{\top} \frac{\sum_{m=1}^{M_j}w_m^{(j)}\boldsymbol{h}_{m}^{(j)}}{||\boldsymbol{o}_j||} = \frac{1}{M_j}\sum_{m=1}^{M_j} \tilde{w}_{m}^{(j)} \cos\theta_{j, 1, m},
\label{eq:decompose}
\end{equation} 
where $\tilde{w}_{m}^{(j)} = w_m^{(j)} M_j \frac{||\boldsymbol{h}_m^{(j)} ||}{||\boldsymbol{o}_j||}$ and $\cos\theta_{j, 1, m} = \widehat{\boldsymbol{c}}_1^{\top}\widehat{\boldsymbol{h}}_{m}^{(j)}$. 
We define $s_{j,m} \equiv \tilde{w}_m^{(j)} \cos\theta_{j, 1, m}$ as the score of the $m$-th segment in the $j$-th trial, which measures the weighted cosine distance between the bona fide class vector $\widehat{\boldsymbol{c}}_1$ and the feature vector $\widehat{\boldsymbol{h}}_m$. Note that $s_{j,m}$ can be larger than one, and the average of $s_{j,m}$ is equal to $s_{j}$. Also note that the decomposition is valid even if $\boldsymbol{o}_j = \mathcal{F}(\sum_{m=1}^{M_j}w_m^{(j)}\boldsymbol{h}_{m}^{(j)})$ where $\mathcal{F}(\cdot)$ is a linear or affine transformation (i.e., a FC layer).

\subsection{CM training for segmental-level detection}
\label{sec:segment} 

In the previous section, we derive segmental scores from an utterance-level detection score. Alternatively, we may train CMs with segment-level labels included in the PartialSpoof database and we may infer a sequence of segment-level scores directly, as Figure~\ref{fig:spfcon_nn} illustrates. 

This segmental CM may use the same LCNN as those in Sec.~\ref{sec:utterance} but with the pooling layer excluded. 
After converting $\boldsymbol{x}_{1:N^{(j)}}\mapsto\boldsymbol{h}_{1:M_j}$ using the LCNN,
the segmental CM computes $\cos\theta_{j, k, m} = \widehat{\boldsymbol{c}}_k^{\top}\widehat{\boldsymbol{h}}_{m}^{(j)}, \forall{m}\in[1, M_j]$ for each segment. Thus, its training loss becomes  
\begin{equation}
\mathcal{L}^{(\text{seg})} = \frac{1}{|\mathcal{D}|}\frac{1}{M_j}\sum_{j=1}^{|\mathcal{D}|}\sum_{m=1}^{M_j}\sum_{k=1}^{C} (\cos\theta_{j,k,m} - \mathds{1}(y_{j,m}=k)) ^ 2,
\end{equation}
where $y_{j,m}$ is the label for the $m$-th segment in the $j$-th trial. During inference, we can directly use $s_{j,m}=\cos\theta_{j,1,m}$ as the segment score for the $m$-th segment.\footnote{If it is necessary to produce a single utterance-level score, one possibility is to use $s_j = \min_{m} {s_{j,m} }$ as the score for the $j$-th trial. This is because a segment with a smaller score is more likely to be spoofed, and a spoofed segment declares a spoofed trial (see step 3 of Sec.~\ref{sec:databse}).} 

\section{Experiments}\label{sec:experiment}
We introduce experimental configurations for our CMs in Sec.~\ref{sec:experiment_conf}. Then, we discuss the performance of utterance- and segmental-level detection in Sec.~\ref{sec:ana_utt} and Sec.~\ref{sec:ana_seg}, respectively. 

\subsection{Experimental configurations}\label{sec:experiment_conf}
All the CMs used linear frequency cepstral coefficients (LFCCs) as input acoustic features. LFCCs were extracted using the same configuration as the ASVspoof 2019 baseline: frame length of 20~ms, frame shift of~10 ms, 512-point FFT, linear filter-bank with 20 channels, and a combination of static, delta, and delta-delta coefficients, thus 60 dimensions for each frame. We did not use any data augmentation, voice activity detection, or feature normalization. 

The LCNN component\footnote{Specifically, from layer {Conv\_1} to {MaxPool\_28}.} in all the CMs was based on the original LCNN-based CM~\cite{lavrentyeva2019stc}. Accordingly, the input LFCCs $\boldsymbol{x}_{1:N^{(j)}}$ are converted into $\boldsymbol{h}_{1:N^{(j)}/16}$ before pooling. In this way, embedding can be extracted every 0.16 seconds. Training  was conducted with the Adam optimizer ($\beta_1=0.9, \beta_2=0.999, \epsilon=10^{-8}$) \cite{kingma2014adam} and a batch size of 64. The learning rate started from $3\times10^{-4}$ and was halved for every 10 epochs. 
The LCNN was trained mutiple times with random initialization. Results shown in the next sections are the mean of six rounds \footnote{More results can be found in arXiv: \scriptsize{https://arxiv.org/abs/2104.02518}}. All results are reproducible using the same random seed and GPU (Nvidia Tesla V100) environment.

Evaluation was conducted using the Equal Error Rate (EER) and minimum tandem detection cost function (min-tDCF)~\cite{kinnunen2018t, kinnunen2020tandem}, both of which are computed following the official routines from ASVspoof 2019. For min-tDCF evaluation, we defined 63,882 spoofed trials for the evaluation set and 22,296 for the development set, based on the new PartialSpoof database. Bona fide (target and non-target) trials are identical to those of the standard ASVspoof 2019 protocols. 

\subsection{Experiments for utterance-level detection} \label{sec:ana_utt}
\subsubsection{Ablation study of LCNN CMs against PartialSpoof}

We first show CM performance on utterance-level detection. Table~\ref{tab:abalation} shows an ablation study of the enhanced LCNN CMs against partially-spoofed audio at the utterance level. All four CMs were trained with utterance-level labels (Sec. \ref{sec:utt_utt}), and detections were also done at the utterance level. We can see that the AP + Bi-LSTM system yields superior performance; thus it was chosen for further cross-database experiments in Section~\ref{sec:crossdata}. 

\begin{table}[t]
\caption{Ablation study of LCNN countermeasures against partially-spoofed audio.}
\label{tab:abalation}
 \vspace*{-2mm}
\centering
\setlength{\tabcolsep}{4pt}
\begin{tabular}{cccccc}
\toprule
\textbf{Pooling} & \textbf{Smoothing} & \multicolumn{2}{c}{\textbf{EER(\%)}}  & \multicolumn{2}{c}{\textbf{min-tDCF}}    \\
\textbf{types}   & \textbf{types}     & \textbf{Dev.} & \textbf{Eval.} & \textbf{Dev.} & \textbf{Eval.} \\
\midrule
AP   &  -                    & 3.90                 & 7.78         & 0.0981      & 0.1972  \\
SAP  &  -                    & 3.90                 & 7.45         & 0.1033      & 0.1887 \\
AP   &  Bi-LSTM              & \textbf{3.68}                 & \textbf{6.19}         & \textbf{0.1003}      & 0.1645 \\
SAP  &  Bi-LSTM              & 3.84        			& 6.23		   & 0.1064      & \textbf{0.1609}    \\
 \bottomrule
 \end{tabular}
\end{table}

\subsubsection{Cross-database investigation for training data mismatch}\label{sec:crossdata}

We investigated how training data mismatch affects CM performance on utterance-level detection. More specifically, we investigated two questions: Can the LCNNs trained using partially(entirely)-spoofed audio detect entirely(partially)-spoofed utterances?
Thus, we trained CMs on the PartialSpoof database and evaluated it on the evaluation set of the ASVspoof 2019 database and vice versa. We also included the matched (in-domain) cases as reference. Results are shown in Table~\ref{tab:cross-database}. 

\begin{table}[t]
\centering
\footnotesize  
\caption{Cross-database study for investigating how training data mismatch affects CM performance. The best architecture (Utterance + AP + Bi-LSTM) in Table \ref{tab:abalation} has been selected.}
 \vspace*{-2mm}
\label{tab:cross-database}
\setlength{\tabcolsep}{4pt}
\begin{tabular}{cccccc}
\toprule
        &       & \multicolumn{2}{c}{\textbf{ASVspoof 2019}}  & \multicolumn{2}{c}{\textbf{PartialSpoof}}     \\
&\textbf{Train} & \textbf{Dev.} & \textbf{Eval.}          & \textbf{Dev.} & \textbf{Eval.}  \\
\midrule
\multirow{2}{*}{\textbf{EER(\%)}} & ASVspoof 2019  & 0.21      & 2.65    & 9.59             & 15.96                        \\
& PartialSpoof  & 4.28      & 5.38    & 3.68             & 6.19                       \\
\midrule
\multirow{2}{*}{\textbf{min-tDCF}} & ASVspoof 2019  & 0.0060      & 0.0640    & 0.1854     & 0.3003                       \\
& PartialSpoof  & 0.1156     & 0.1713    & 0.1003             & 0.1645                        \\
\bottomrule
\end{tabular}
\vspace{-3mm}
\end{table}
By comparing results in each set of the ASVspoof 2019 dataset with counterparts of the PartialSpoof dataset, we can first see that the EER for the ASVspoof 2019 dataset is always lower than that for the PartialSpoof dataset on both the sets, showing that partially-spoofed data is more difficult to detect than fully-spoofed data. 

Next we can see that when the ASVspoof 2019-trained model is evaluated on the PartialSpoof database, performance degrades significantly; the EER increases from 0.21\% to 9.59\% and from 2.65\% to 15.96\% for development and evaluation sets, respectively. On the other hand, the model trained on the partially-spoofed database is relatively robust and shows a stable EER when evaluated on both databases. The models trained on fully-spoofed utterances appear to lack generalization ability and thus overfit to the fully-spoofed case. 


\begin{figure}[t]
  \centering
    \includegraphics[width=1.0\linewidth]{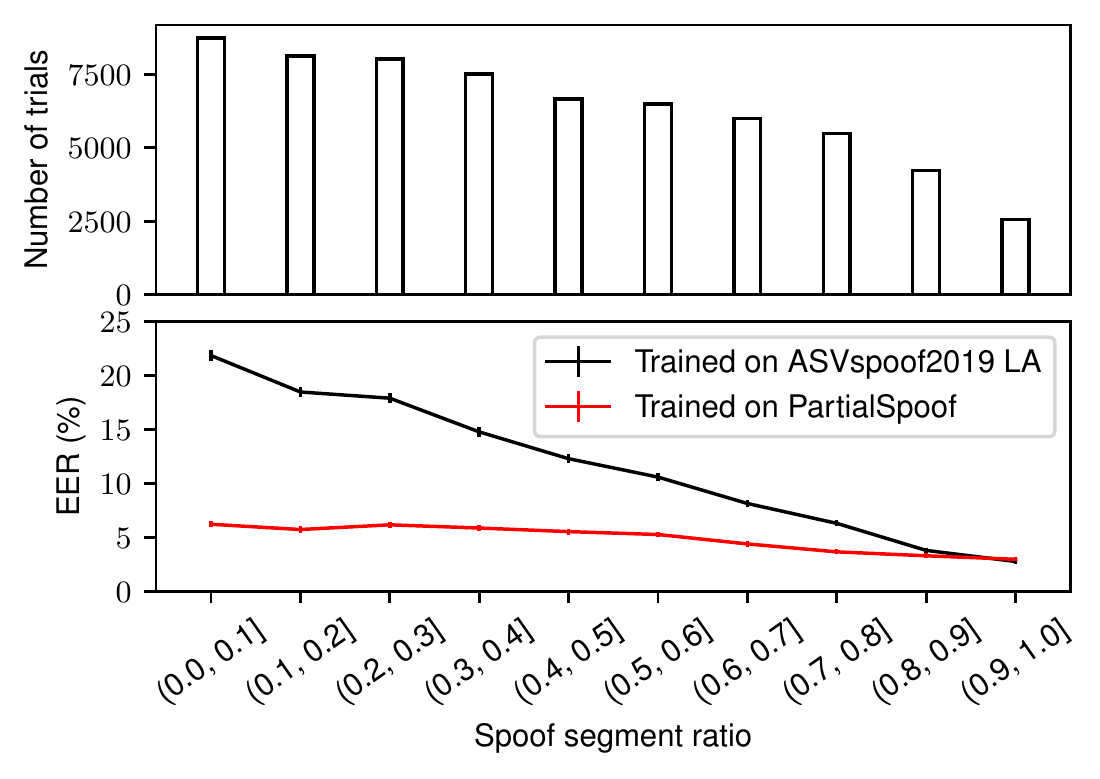}
  \vspace{-7mm}
  \caption{Break-down of results of the cross-database study. (a) Histogram of number of trials having different spoof segment ratios. The ratio was quantized for visualization purposes. (b) EERs for each of the quantized spoof segment ratio classes with confidence intervals at a significance level of 5\% \cite{bengio2004statistical}. The best LCNN network (AP + Bi-LSTM) in Table \ref{tab:abalation} has been selected.}
  \label{fig:eval-res}
\end{figure}

\subsubsection{Analysis based on spoof segment ratios}

Since the PartialSpoof database was constructed from the random replacement of speech segments, the total duration occupied by multiple injected spoofed segments within in an utterance can vary.  We refer to the ratio of the total duration of spoofed segments within an entire length of audio as the ``spoof segment ratio'' and further investigate how the detection performance changes according to the spoof segment ratio of the trials. For this analysis, we quantized the spoof segment ratio into 10 bins and computed the EER for each bin separately. 

\begin{figure}[t]
  \centering
    \includegraphics[width=1.0\linewidth]{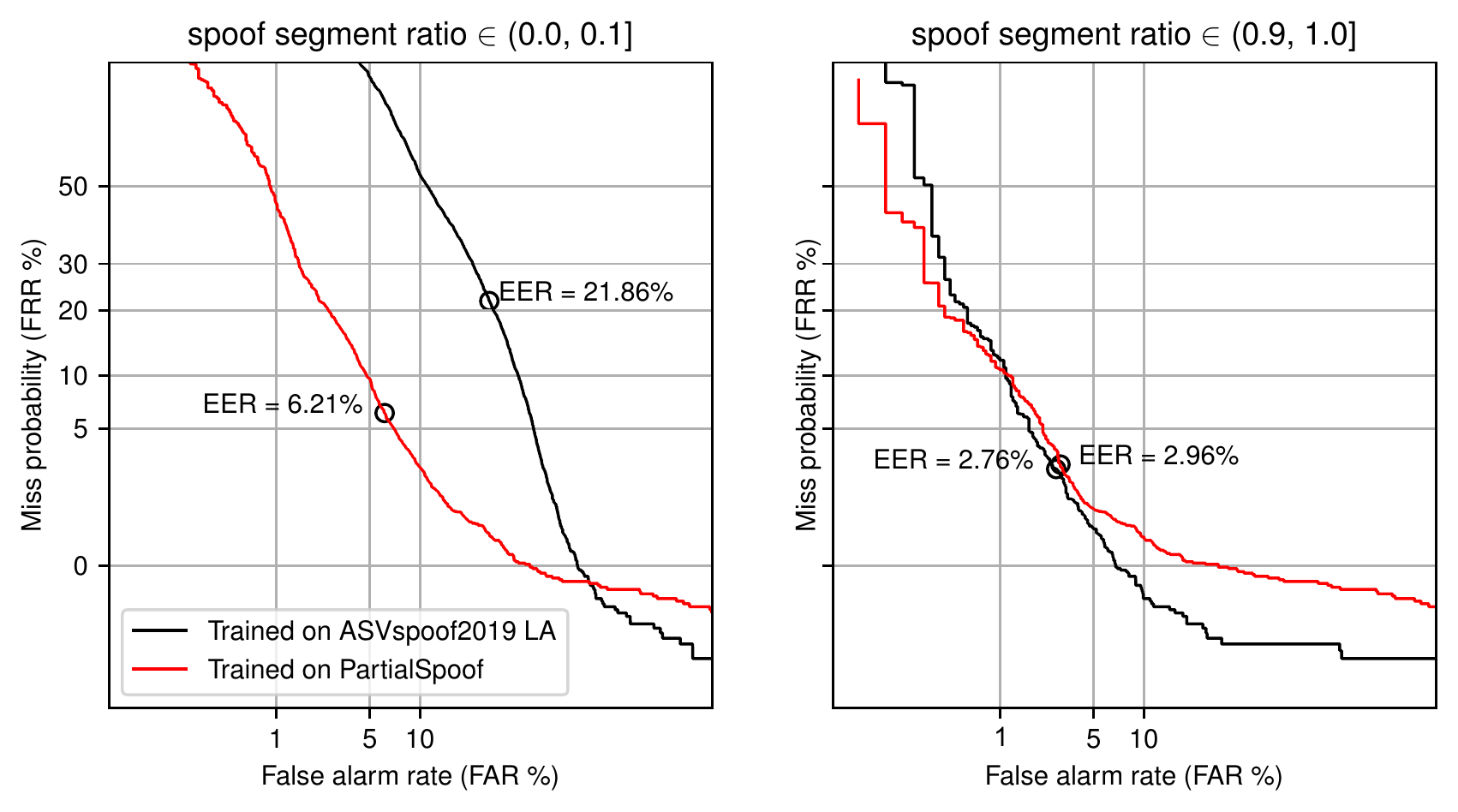}
  \vspace{-6mm}
  \caption{Comparison of DET curves in the eval.\ set of the PartialSpoof database at areas of different spoof segment ratios.}
  \label{fig:eval-det}
  \vspace{-5mm}
\end{figure}

Figure~\ref{fig:eval-res} shows the relationship between the quantized spoof segment ratio and performance. The top histogram presents the number of trials included in each bin. We evaluate EER for each bin using its corresponding spoofed trials and all bona fide trials and then obtain the bottom two curves in the lower plot. From this figure, we can confirm that the spoof segment ratio has a significant impact upon the CM performance. More specifically, we see that when the CM model is trained using fully-spoofed audio, as expected, the EER degrades with decreases in the spoof segment ratio. On the other hand, the CM model trained using partially spoofed audio is robust to changes in the spoof segment ratio. Even if the ratio changes, EER values do not change significantly. 

Figure~\ref{fig:eval-det} shows the DET curves at the smallest and largest spoof segment ratios of the PartialSpoof evaluation set. From Figure~\ref{fig:eval-res} and Figure~\ref{fig:eval-det}, we can reconfirm that the reliability of countermeasures trained to detect fully-spoofed data degrade substantially when tested with partially-spoofed data.

\subsection{Experiments for segmental-level detection}\label{sec:ana_seg}

\begin{table}[t]
\caption{Comparison of segmental detection performance of CMs trained using segmental or utterance-level labels.}
\label{tab:segment-detection}
 \vspace*{-2mm}
\centering
\setlength{\tabcolsep}{4pt}
\begin{tabular}{cccccc}
\toprule
\textbf{Train}  &  \textbf{Pooling} & \textbf{Smoothing} & \multicolumn{2}{c}{\textbf{EER(\%)}}    \\
\textbf{labels} & \textbf{types}   & \textbf{types}     & \textbf{Dev.} & \textbf{Eval.}  \\
\midrule
Utterance       & AP   &  Bi-LSTM                       & 37.02       &  40.20   \\
Segment         & -    &  Bi-LSTM                       & 6.81        &  16.21   \\
 \bottomrule
 \end{tabular}
 \vspace{-4mm}
\end{table}

Next we focus on segmental-level detection. This is challenging because an individual segment can be very short. The main focus is a CM trained using the segmental labels instead of utterance labels as described in Sec.~\ref{sec:segment}. For reference, we compare it with another CM trained using utterance labels and their segmental scores derived from an utterance-level score in the manner described in  Sec.~\ref{sec:utt_seg}. 

Table \ref{tab:segment-detection} shows the segmental detection results. Not surprisingly, the CM trained using the segmental labels is better than the one trained using utterance-level labels. This shows that the segmental labels included in the PartialSpoof database are useful for segmental detection, and that segment detection is feasible. But we can also see that this is a more challenging task than the utterance-level detection in the previous section, and the CMs have obvious room for further improvement. 

\section{Conclusions}\label{sec:conclusion}

To answer the original question: \textit{`can we detect partially spoofed audio?'}, we built a new PartialSpoof database consisting of bona fide and partially-spoofed utterances based on ASVspoof 2019. Since PartialSpoof audio is composed of bona fide and spoofed segments(s), it can be trained and evaluated on both utterance- and segmental- level labels. For utterance-level detection, cross-database analyses on partially- and fully-spoofed data were conducted to investigate how data mismatch affects CM performance. We also carried out a more challenging segmental detection task to see whether CMs can spot short spoofed segments included in an utterance.

Generally, both utterance- and segmental-level detection on PartialSpoof are more challenging than on the fully-spoofed database. The reliability of countermeasures trained to detect fully-spoofed data was also found to degrade substantially when tested with partially-spoofed data, while training on partially-spoofed data led to stable performance when evaluating on both fully- and partially-spoofed utterances.

Future studies are needed to understand the data mismatch problem deeply. Furthermore, random segment selection and concatenation using cross-correlation may not be the best way to build a partially-spoofed database. Linguistic information, contextual information, and rhythm can be lost during this process. Further exploration of more appropriate databases and more robust CMs with higher precision are needed.

\vspace{1.5mm}
\noindent
\textbf{Acknowledgements}
{\footnotesize Thanks Dr.\ Ville Vestman from University of Eastern Finland for sharing an ASV model for min-tDCF evaluation in this paper. This study was partially supported by the Japanese-French joint national VoicePersonae project supported by JST CREST (JPMJCR18A6) and the ANR (ANR-18-JSTS-0001), JST CREST Grants (JPMJCR20D3), MEXT KAKENHI Grants (16H06302, 18H04120, 18H04112, 18KT0051), Japan, and Google AI for Japan program. }

\bibliographystyle{IEEEtran}
\bibliography{main}

\begin{thebibliography}{10}
\providecommand{\url}[1]{#1}
\csname url@samestyle\endcsname
\providecommand{\newblock}{\relax}
\providecommand{\bibinfo}[2]{#2}
\providecommand{\BIBentrySTDinterwordspacing}{\spaceskip=0pt\relax}
\providecommand{\BIBentryALTinterwordstretchfactor}{4}
\providecommand{\BIBentryALTinterwordspacing}{\spaceskip=\fontdimen2\font plus
\BIBentryALTinterwordstretchfactor\fontdimen3\font minus
  \fontdimen4\font\relax}
\providecommand{\BIBforeignlanguage}[2]{{%
\expandafter\ifx\csname l@#1\endcsname\relax
\typeout{** WARNING: IEEEtran.bst: No hyphenation pattern has been}%
\typeout{** loaded for the language `#1'. Using the pattern for}%
\typeout{** the default language instead.}%
\else
\language=\csname l@#1\endcsname
\fi
#2}}
\providecommand{\BIBdecl}{\relax}
\BIBdecl

\bibitem{jain2006biometrics}
A.~K. Jain, A.~Ross, and S.~Pankanti, ``Biometrics: a tool for information
  security,'' \emph{IEEE transactions on information forensics and security},
  vol.~1, no.~2, pp. 125--143, 2006.

\bibitem{Wu2014}
Z.~Wu, T.~Kinnunen, N.~Evans, J.~Yamagishi, C.~Hanil{\c{c}}i, M.~Sahidullah,
  and A.~Sizov, ``{ASVspoof 2015: the First Automatic Speaker Verification
  Spoofing and Countermeasures Challenge},'' in \emph{Proc. Interspeech}, 2015,
  pp. 2037--2041.

\bibitem{Kinnunen2017}
T.~Kinnunen, M.~Sahidullah, H.~Delgado, M.~Todisco, N.~Evans, J.~Yamagishi, and
  K.~A. Lee, ``{The ASVspoof 2017 Challenge: Assessing the Limits of Replay
  Spoofing Attack Detection},'' in \emph{Proc. Interspeech}, 2017, pp. 2--6.

\bibitem{Nautsch2021}
A.~Nautsch, X.~Wang, N.~Evans, T.~Kinnunen, V.~Vestman, M.~Todisco, H.~Delgado,
  M.~Sahidullah, J.~Yamagishi, and K.~A. Lee, ``{ASVspoof 2019: spoofing
  countermeasures for the detection of synthesized, converted and replayed
  speech},'' \emph{IEEE Transactions on Biometrics, Behavior, and Identity
  Science}, 2021.

\bibitem{lavrentyeva2019stc}
G.~Lavrentyeva, S.~Novoselov, A.~Tseren, M.~Volkova, A.~Gorlanov, and
  A.~Kozlov, ``{STC Antispoofing Systems for the ASVspoof2019 Challenge},'' in
  \emph{Proc. Interspeech}, 2019, pp. 1033--1037.

\bibitem{Chen2020Odyssey}
T.~Chen, A.~Kumar, P.~Nagarsheth, G.~Sivaraman, and E.~Khoury,
  ``{Generalization of Audio Deepfake Detection},'' in \emph{Proc. Odyssey},
  2020, pp. 132--137.

\bibitem{Chettri2019}
B.~Chettri, D.~Stoller, V.~Morfi, M.~A.~M. Ram{\'{i}}rez, E.~Benetos, and B.~L.
  Sturm, ``{Ensemble Models for Spoofing Detection in Automatic Speaker
  Verification},'' in \emph{Proc. Interspeech}, 2019, pp. 1018--1022.

\bibitem{tak2020end}
H.~Tak, J.~Patino, M.~Todisco, A.~Nautsch, N.~Evans, and A.~Larcher,
  ``End-to-end anti-spoofing with rawnet2,'' in \emph{Proc. ICASSP}, 2021, pp.
  6369--6373.

\bibitem{wang2021comparative}
X.~Wang and J.~Yamagishi, ``A comparative study on recent neural spoofing
  countermeasures for synthetic speech detection,'' in \emph{Proc.
  Interspeech}, 2021 (to appear).

\bibitem{Wang2020data}
X.~Wang, J.~Yamagishi, M.~Todisco, H.~Delgado, A.~Nautsch, N.~Evans,
  M.~Sahidullah, V.~Vestman, T.~Kinnunen, K.~A. Lee, L.~Juvela, P.~Alku, Y.-H.
  Peng, H.-T. Hwang, Y.~Tsao, H.-M. Wang, S.~L. Maguer, M.~Becker,
  F.~Henderson, R.~Clark, Y.~Zhang, Q.~Wang, Y.~Jia, K.~Onuma, K.~Mushika,
  T.~Kaneda, Y.~Jiang, L.-J. Liu, Y.-C. Wu, W.-C. Huang, T.~Toda, K.~Tanaka,
  H.~Kameoka, I.~Steiner, D.~Matrouf, J.-F. Bonastre, A.~Govender, S.~Ronanki,
  J.-X. Zhang, and Z.-H. Ling, ``Asvspoof 2019: A large-scale public database
  of synthesized, converted and replayed speech,'' \emph{Computer Speech and
  Language}, vol.~64, p. 101114, 2020.

\bibitem{povey2011kaldi}
D.~Povey, A.~Ghoshal, G.~Boulianne, L.~Burget, O.~Glembek, N.~Goel,
  M.~Hannemann, P.~Motlicek, Y.~Qian, P.~Schwarz \emph{et~al.}, ``The {Kaldi}
  speech recognition toolkit,'' in \emph{Proc. ASRU}, 2011.

\bibitem{kinnunen2010overview}
T.~Kinnunen and H.~Li, ``{An overview of text-independent speaker recognition:
  From features to supervectors},'' \emph{Speech communication}, vol.~52,
  no.~1, pp. 12--40, 2010.

\bibitem{Lavechin-sad-dihard}
M.~Lavechin, M.-P. Gill, R.~Bousbib, H.~Bredin, and L.~{Paola Garcia-Perera},
  ``{End-to-end Domain-Adversarial Voice Activity Detection},'' in \emph{Proc.
  Interspeech}, 2020, pp. 3685--3689.

\bibitem{pyannote}
H.~{Bredin}, R.~{Yin}, J.~M. {Coria}, G.~{Gelly}, P.~{Korshunov},
  M.~{Lavechin}, D.~{Fustes}, H.~{Titeux}, W.~{Bouaziz}, and M.~{Gill},
  ``Pyannote.audio: Neural building blocks for speaker diarization,'' in
  \emph{Proc. ICASSP}, 2020, pp. 7124--7128.

\bibitem{pyannote.metrics}
H.~Bredin, ``{pyannote.metrics: a toolkit for reproducible evaluation,
  diagnostic, and error analysis of speaker diarization systems},'' in
  \emph{{Proc. Interspeech}}, Stockholm, Sweden, August 2017.

\bibitem{Zue1990timit}
V.~Zue, S.~Seneff, and J.~Glass, ``{Speech database development at MIT: Timit
  and beyond},'' \emph{Speech Communication}, vol.~9, no.~4, pp. 351--356, aug
  1990.

\bibitem{sv56}
{International Telecommunication Union}, Recommendation {G.191}: Software Tools
  and Audio Coding Standardization, Nov 11 2005.

\bibitem{wu2018light}
X.~Wu, R.~He, Z.~Sun, and T.~Tan, ``{A light cnn for deep face representation
  with noisy labels},'' \emph{IEEE Transactions on Information Forensics and
  Security}, vol.~13, no.~11, pp. 2884--2896, 2018.

\bibitem{Zhu2018}
Y.~Zhu, T.~Ko, D.~Snyder, B.~Mak, and D.~Povey, ``{Self-Attentive Speaker
  Embeddings for Text-Independent Speaker Verification},'' in \emph{Proc.
  Interspeech}, 2018, pp. 3573--3577.

\bibitem{kingma2014adam}
D.~P. Kingma and J.~Ba, ``{Adam: A method for stochastic optimization},'' in
  \emph{Proc. ICLR}, 2014.

\bibitem{kinnunen2018t}
T.~Kinnunen, K.~A. Lee, H.~Delgado, N.~Evans, M.~Todisco, M.~Sahidullah,
  J.~Yamagishi, and D.~A. Reynolds, ``{t-DCF: a detection cost function for the
  tandem assessment of spoofing countermeasures and automatic speaker
  verification},'' in \emph{Proc. Odyssey}, 2018, pp. 312--319.

\bibitem{kinnunen2020tandem}
T.~Kinnunen, H.~Delgado, N.~Evans, K.~A. Lee, V.~Vestman, A.~Nautsch,
  M.~Todisco, X.~Wang, M.~Sahidullah, J.~Yamagishi, and D.~A. Reynolds,
  ``{Tandem Assessment of Spoofing Countermeasures and Automatic Speaker
  Verification: Fundamentals},'' \emph{IEEE/ACM Transactions on Audio, Speech,
  and Language Processing}, vol.~28, pp. 2195--2210, 2020.

\bibitem{bengio2004statistical}
S.~Bengio and J.~Mari{\'{e}}thoz, ``{A statistical significance test for person
  authentication},'' in \emph{Proc. Odyssey}, 2004.

\end{thebibliography}

\newpage
\clearpage
\onecolumn

\begin{appendices}
\section{Appendix}

\subsection{Details of PartialSpoof Database\protect\footnote{Samples can be found at https://nii-yamagishilab.github.io/zlin-demo/IS2021/index.html}}

\subsubsection{Database collection}
\begin{figure}[!hbp]
  \centering
    \includegraphics[width=1.0\linewidth]{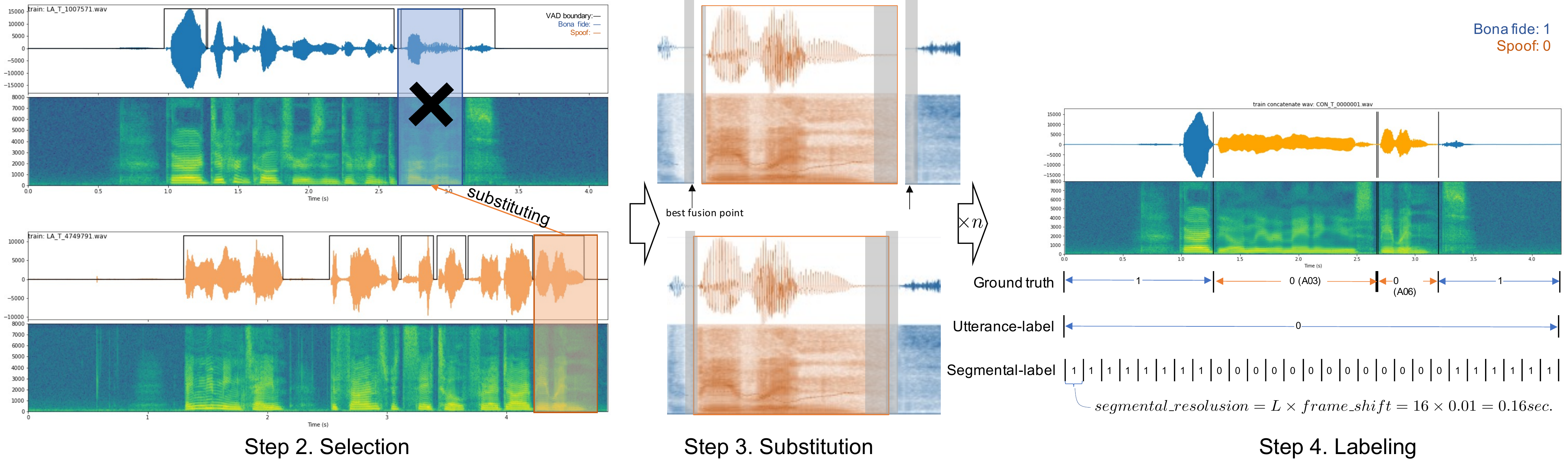}
  \vspace{-6mm}
  \caption{Procedure of data collection. (Example of CON\_T\_0000001.wav)}
  \label{fig:data_collection}
  \vspace{-5mm}
\end{figure}

Figure \ref{fig:data_collection} shows the generated procedure for one partially-spoofed audio (Step 2-4 in Section \ref{sec:databse}) . In this case (CON\_T\_000001.wav), we substitute the bona fide segment(s) LA\_T\_1007571$[42160: 49440]$\footnote{in sample point} and LA\_T\_1007571$[20400: 41942]$ by using spoofed segment(s) LA\_T\_4749791$[67600: 76000]$ and LA\_T\_8410615$[96640: 119200]$, respectively. For the whole PartialSpoof database, Step 2-4 will be repeated until we get the same number of spoofed trials as the ASVspoof 2019 database.

\subsubsection{Statistics of spoof trails in PartialSpoof dataset}

In the PartialSpoof dataset, bona fide trails are identical to the ASVspoof 2019 LA database. Spoof trails are partially-spoofed, generated from ASVspoof 2019 LA database. The statistics of spoofed trails in the PartialSpoof dataset can be seen in Table \ref{tab:data_stat}
\begin{table}[h]
\caption{Statistics of spoofed trails in PartialSpoof dataset. Entries that have three values are reported as min/mean/max.    \# Num: Total number in the PartialSpoof dataset. \# Tol Duration (sec.): Total duration in the dataset. Spoof System: Maximum number of different spoof systems within utterance.  Duration (sec.):
Length of each audio in seconds. spoof segment ratio \%: Percentage of audio time that is spoofed in an audio. (Only spoof information here)}
\label{tab:data_stat}
\begin{center}
\begin{tabular}{cccccc}
\toprule
\textbf{Set}   & \textbf{\# Num} & \textbf{\# Tol Duration (sec.)} & \textbf{Spoof System} & \textbf{Durations (sec.)} & \textbf{Spoof Segment Ratio (\%)} \\
\midrule
\textbf{Train} & 22800           & 78577.64                     & 6                     & 0.60/3.45/21.02           & 0.65/47.63/99.80                          \\
\textbf{Dev.}  & 22296           & 78705.49                     & 6                     & 0.62/3.53/15.34           & 0.72/46.78/99.77                          \\
\textbf{Eval.} & 63882           & 218667.47                    & 9                     & 0.48/3.42/18.20           & 0.23/41.97/99.81                             \\
\bottomrule
\end{tabular}
\end{center}
\vspace{-5mm}
\end{table}


\subsection{Multiple random initialization for Section \ref{sec:experiment}}
\begin{normalsize}
We mentioned in Section \ref{sec:experiment} that all CMs were trained mutiple times with random initialization where $random\_seed = \{10^{k-1}, k=1, \cdots, 6\}$. Final results are the mean of them. Results for each round are shown in this section in detail. Table \ref{tab:ablation_6seed_eer}, \ref{tab:ablation_6seed_mintdcf} shows details for ablation study. Table \ref{tab:cross_6seed_asvspf19_eval}, \ref{tab:cross_6seed_partial_eval} displays details for cross-database study. Table \ref{tab:segmental_pred_6seed} demonstrates details of segmental detection of Section \ref{sec:ana_seg}.
\end{normalsize}

\subsubsection{Ablation study}

\begin{table}[!h] 
\footnotesize
\caption{Ablation study of LCNN countermeasures against partially-spoofed audio. EER (\%) on PartialSpoof dev. and eval. set. (Details of Table \ref{tab:abalation})}
\label{tab:ablation_6seed_eer}
\resizebox{\textwidth}{!}{
\begin{tabular}{cccccccccccccccc}

\toprule
\textbf{Pooling} & \textbf{Smoothing} & \multicolumn{7}{c}{\textbf{Dev.}}                                                               & \multicolumn{7}{c}{\textbf{Eval.}}                                                              \\
\cline{3-16}
\textbf{types}   & \textbf{types}     &  \textbf{1} & \textbf{10} & \textbf{100} & \textbf{1000} & \textbf{10000} & \textbf{100000} & \textbf{Mean} & \textbf{1} & \textbf{10} & \textbf{100} & \textbf{1000} & \textbf{10000} & \textbf{100000} & \textbf{Mean} \\
\midrule
AP               & -                  & 3.54        & 4.40        & 3.50         & 4.12          & 3.93           & 3.93            & 3.90          & 7.83       & 7.59        & 7.88         & 8.16          & 7.38           & 7.85            & 7.78          \\
SAP              & -                  & 4.04        & 4.16        & 3.81         & 3.93          & 3.76           & 3.68            & 3.90          & 7.41       & 7.68        & 7.44         & 7.42          & 7.84           & 6.94            & 7.45          \\
AP               & LSTM               & 3.73        & 3.57        & 3.80         & 2.68          & 2.80           & 5.49            & 3.68          & 6.45       & 6.31        & 6.51         & 6.01          & 6.01           & 5.83            & 6.19          \\
SAP              & LSTM               & 3.37        & 3.69        & 3.53         & 4.59          & 4.64           & 3.23            & 3.84          & 6.06       & 6.54        & 5.98         & 6.71          & 6.20           & 5.87            & 6.23   \\
\bottomrule
\end{tabular}
}
\end{table}

\begin{table}[!ht]
\footnotesize
\caption{Ablation study of LCNN countermeasures against partially-spoofed audio. min-tDCF on PartialSpoof dev. and eval. set. (Details of Table \ref{tab:abalation}) }
\label{tab:ablation_6seed_mintdcf}
\resizebox{\textwidth}{!}{
\begin{tabular}{cccccccccccccccc}
\toprule
\textbf{Pooling} & \textbf{Smoothing} & \multicolumn{7}{c}{ \textbf{Dev.}}                                                                   & \multicolumn{7}{c}{\textbf{Eval}}                                                              \\
\cline{3-16}
\textbf{types}   & \textbf{types}     &  \textbf{1} & \textbf{10}     & \textbf{100} & \textbf{1000} & \textbf{10000} & \textbf{100000} & \textbf{Mean} & \textbf{1} & \textbf{10} & \textbf{100} & \textbf{1000} & \textbf{10000} & \textbf{100000} & \textbf{Mean} \\
\midrule
\textbf{AP}      & \textbf{-}         & 0.0884     & 0.1099          & 0.0882       & 0.1023        & 0.1011         & 0.0991          & 0.0981        & 0.1937     & 0.1917      & 0.2020       & 0.2120        & 0.1849         & 0.1988          & 0.1972        \\
\textbf{SAP}     & \textbf{-}         & 0.1074     & 0.1163          & 0.1009       & 0.1055        & 0.0972         & 0.0929          & 0.1033        & 0.1827     & 0.1950      & 0.1903       & 0.1871        & 0.1991         & 0.1780          & 0.1887        \\
\textbf{AP}      & \textbf{LSTM}      & 0.1072     & 0.1027          & 0.1098       & 0.0778        & 0.0688         & 0.1358          & 0.1003        & 0.1834     & 0.1644      & 0.1760       & 0.1558        & 0.1466         & 0.1611          & 0.1645        \\
\textbf{SAP}     & \textbf{LSTM}      & 0.0975     & 0.1061          & 0.0978       & 0.1282        & 0.1224         & 0.0865          & 0.1064        & 0.1483     & 0.1607      & 0.1554       & 0.1839        & 0.1736         & 0.1435          & 0.1609 \\  
\bottomrule
\end{tabular}}
\end{table}

\subsubsection{Cross-database study}

\begin{table}[!h]
\small
\caption{Cross-database study based on Utterance + AP + Bi-LSTM. Evaluated on ASVspoof 2019 database. (Details of Table \ref{tab:cross-database})}
\label{tab:cross_6seed_asvspf19_eval}
\resizebox{\textwidth}{!}{
\begin{tabular}{cccccccccccccccc}
\toprule
                                    &                                  & \multicolumn{7}{c}{{ \textbf{Dev.}}}                                                                            & \multicolumn{7}{c}{\textbf{Eval.}}                                                                         \\
\cline{3-16}
\multirow{-2}{*}{\textbf{}}         & \multirow{-2}{*}{\textbf{Train}} &  \textbf{1} & \textbf{10}   & \textbf{100} & \textbf{1000} & \textbf{10000} & \textbf{100000} & \textbf{Mean} & \textbf{1} & \textbf{10} & \textbf{100} & \textbf{1000} & \textbf{10000} & \textbf{100000} & \textbf{Mean} \\
\midrule
                                    & \textbf{ASVspoof 2019}              & 0.24                     & 0.20 & 0.12         & 0.31          & 0.20           & 0.20            & 0.21          & 2.64       & 2.79        & 2.31         & 2.30          & 3.06           & 2.81            & 2.65          \\
\multirow{-2}{*}{\textbf{EER (\%)}} & \textbf{PartialSpoof}            & 5.22                              & 5.77          & 4.52         & 2.94          & 0.94           & 6.32            & 4.28          & 6.12       & 5.91        & 5.59         & 4.76          & 4.88           & 5.03            & 5.38          \\
\midrule                                    
                                    & \textbf{ASVspoof 2019}              & 0.0064                            & 0.0064        & 0.0033       & 0.0080        & 0.0056         & 0.0064          & 0.0060        & 0.0689     & 0.0622      & 0.0611       & 0.0617        & 0.0645         & 0.0659          & 0.0640        \\
\multirow{-2}{*}{\textbf{min-tDCF}} & \textbf{PartialSpoof}            & 0.1357                            & 0.1493        & 0.1393       & 0.0771        & 0.0315         & 0.1607          & 0.1156        & 0.2057     & 0.1909      & 0.1889       & 0.1533        & 0.1187         & 0.1703          & 0.1713       \\
\bottomrule
\end{tabular}
}
\end{table}

\begin{table}[!h]
\footnotesize
\caption{Cross-database study based on Utterance + AP + Bi-LSTM. Evaluated on PartialSpoof database. (Details of Table \ref{tab:cross-database})}
\label{tab:cross_6seed_partial_eval}
\resizebox{\textwidth}{!}{
\begin{tabular}{cccccccccccccccc}
\toprule
                                    &                                  & \multicolumn{7}{c}{{ \textbf{Dev.}}}                                                                             & \multicolumn{7}{c}{\textbf{Eval.}}                                                                         \\
\cline{3-16}
\multirow{-2}{*}{\textbf{}}         & \multirow{-2}{*}{\textbf{Train}} & \textbf{1} & \textbf{10}    & \textbf{100} & \textbf{1000} & \textbf{10000} & \textbf{100000} & \textbf{Mean} & \textbf{1} & \textbf{10} & \textbf{100} & \textbf{1000} & \textbf{10000} & \textbf{100000} & \textbf{Mean} \\
\midrule
                                    & \textbf{ASVspoof 2019}              & 8.83                     & 10.04 & 10.56        & 7.97          & 10.36          & 9.77            & 9.59          & 14.70      & 16.93       & 16.68        & 14.20         & 17.91          & 15.33           & 15.96         \\
\multirow{-2}{*}{\textbf{EER (\%)}} & \textbf{PartialSpoof}            & 3.73        & 3.57        & 3.80         & 2.68          & 2.80           & 5.49            & 3.68          & 6.45       & 6.31        & 6.51         & 6.01          & 6.01           & 5.83            & 6.19          \\
\midrule
                                    & \textbf{ASVspoof 2019}              & 0.1687                            & 0.2004         & 0.1956       & 0.1656        & 0.1860         & 0.1960          & 0.1854        & 0.2872     & 0.3235      & 0.3100       & 0.2707        & 0.3073         & 0.3034          & 0.3003        \\
\multirow{-2}{*}{\textbf{min-tDCF}} & \textbf{PartialSpoof}            & 0.1072     & 0.1027          & 0.1098       & 0.0778        & 0.0688         & 0.1358          & 0.1003        & 0.1834     & 0.1644      & 0.1760       & 0.1558        & 0.1466         & 0.1611          & 0.1645        \\
\bottomrule
\end{tabular}
}  
\end{table}



\subsubsection{Segmental-level detection}

\begin{table}[!h]
\caption{Comparison of segmental detection performance of CMs trained using segmental or utterance-level labels. (Details of Table \ref{tab:segment-detection})}
\label{tab:segmental_pred_6seed}
\resizebox{\textwidth}{!}{
\begin{tabular}{ccccccccccccccccc}
\toprule
\textbf{Train}     & \textbf{Pooling} & \textbf{Smoothing} & \multicolumn{7}{c}{\textbf{Dev.}}                                                                          & \multicolumn{7}{c}{\textbf{Eval.}}                                                                         \\
\cline{4-17}
\textbf{labels}    & \textbf{types}   & \textbf{types}     & \textbf{1} & \textbf{10} & \textbf{100} & \textbf{1000} & \textbf{10000} & \textbf{100000} & \textbf{Mean} & \textbf{1} & \textbf{10} & \textbf{100} & \textbf{1000} & \textbf{10000} & \textbf{100000} & \textbf{Mean} \\
\midrule
\textbf{Utterance} & AP               & Bi-LSTM            & 38.19      & 37.83       & 34.84        & 38.68         & 34.26          & 38.28           & 37.02         & 41.33      & 40.23       & 39.90        & 41.39         & 38.08          & 40.30           & 40.20         \\
\textbf{Segment}            & -                & Bi-LSTM            & 6.59	& 7.07	& 6.92	& 6.53	& 6.96	& 6.79	& 6.81          & 16.53      & 15.93       & 16.00        & 15.59         & 16.60          & 16.62           & 16.21 \\
\bottomrule
\end{tabular}
}
\vspace{-5mm}
\end{table}

\end{appendices}


\end{document}